\documentclass[aps,twocolumn,showlabels,amsmath,amssymb,pre,superscriptaddress, floatfix, colors]{revtex4-1}

\usepackage{amssymb,amsmath,amsfonts}
\usepackage{graphicx,color}
\usepackage[english]{babel}
\usepackage{tabularx}
\usepackage{etoolbox}
\usepackage{graphicx}
\usepackage{setspace}
\usepackage{gensymb}
\usepackage{epstopdf}
\usepackage{mathrsfs}

\usepackage[normalem]{ulem}
\newcommand\ro{\bgroup\markoverwith
{\textcolor{red}{\rule[.3ex]{2pt}{1.5pt}}}\ULon}

\newcommand{\beginsupplement}{
        \setcounter{table}{0}
        \renewcommand{\thetable}{S\arabic{table}}
        \setcounter{figure}{0}
        \renewcommand{\figurename}{Supplementary Figure}
     }

\begin{document}

\title{Intercellular communication induces glycolytic synchronization waves between individually oscillating cells\\ \vspace{1em}}

\author{Martin Mojica-Benavides}
\affiliation{Department of Physics, University of Gothenburg, SE-41296 Gothenburg, Sweden}

\author{David D. van Niekerk}
\affiliation{Department of Biochemistry, Stellenbosch University, Matieland 7602, South Africa}

\author{Mite Mijalkov}
\affiliation{Department of Neurobiology, Care Sciences and Society, Karolinska Institutet, 171 77 Stockholm, Sweden}

\author{Jacky L. Snoep} 
\affiliation{Department of Biochemistry, Stellenbosch University, Matieland 7602, South Africa}
\affiliation{Molecular Cell Physiology, Vrije Universiteit Amsterdam, 1081 HV Amsterdam, The Netherlands}

\author{Bernhard Mehlig}
\affiliation{Department of Physics, University of Gothenburg, SE-41296 Gothenburg, Sweden}

\author{Giovanni Volpe}
\affiliation{Department of Physics, University of Gothenburg, SE-41296 Gothenburg, Sweden}

\author{Mattias Goks\"or}
\thanks{These two authors contributed equally. \\caroline.adiels@physics.gu.se}
\affiliation{Department of Physics, University of Gothenburg, SE-41296 Gothenburg, Sweden}

\author{Caroline B. Adiels} 
\thanks{These two authors contributed equally. \\caroline.adiels@physics.gu.se}
\affiliation{Department of Physics, University of Gothenburg, SE-41296 Gothenburg, Sweden}

\date{\today}

\maketitle
\textbf{
Metabolic oscillations in single cells underlie the mechanisms behind cell synchronization and cell-cell communication.
For example, glycolytic oscillations mediated by biochemical communication between cells may synchronize the pulsatile insulin secretion by pancreatic tissue, and a link between glycolytic synchronization anomalies and type-2 diabetes has been hypotesized.
Cultures of yeast cells have provided an ideal model system to study synchronization and propagation waves of glycolytic oscillations in large populations.
However, the mechanism by which synchronization occurs at individual cell level is still an open question due to experimental limitations in sensitive and specific handling of single cells.
Here, we show how the coupling of intercellular diffusion with the phase regulation of individual oscillating cells induces glycolytic synchronization waves. We directly measure the single-cell metabolic responses from yeast cells in a microfluidic environment and characterize a discretized cell-cell communication using graph theory.
We corroborate our findings with simulations based on a detailed kinetic model for individual yeast cells.
These findings can provide insight into the roles cellular synchronization play in biomedical applications, such as insulin secretion regulation at the cellular level.} \\

Early studies on glycolytic oscillations detected in population measurements proposed acetaldehyde (ACA) as the chemical mediator between oscillating yeast cells \cite{ACA1}. ACA is a metabolite  either produced by the cells themselves \cite{ACA2} or externally supplied \cite{ACA4}. By local addition of glucose (GLC), macroscopic glycoytic synchronization waves can be induced \cite{waves0, waves1, waves2} but not resolved at the single-cell level. Only recently, single-cell analysis has been achieved by using fixed cells on coated microscope slides \cite{single2, synchyeast1} and in alginate microparticles \cite{yeastalginate}. While these approaches manage to obtain biochemical information at the single-cell level, they allow limited control on the environment to characterize the interaction between the cells.
Recently, we have used microfluidics to precisely control the flow fields, and chemical concentrations surrounding yeast cells. This has permitted us to externally entrain the oscillations of single yeast cells by the periodic injection of ACA or a respiration inhibitor i.e., cyanide (CN$^-$) \cite{single5, Dawie2019}. The flow present in these microfluidic systems removes cell secretions, including the ACA mediator required to achieve cell-cell communication. Such an approach prevents studying the process leading from individual cell oscillations to synchronization.

Here, we show how glycolytic synchronization waves are induced by intercellular communication between individual cells and the emergence of coupled subpopulation clusters. We address the process by which the diffusion and reaction of intercellular chemicals couple the oscillation phases of neighboring cells. To do this, we implement a custom-designed diffusion-limited microfluidic device to host a mesoscopic cell culture. While controlling the extracellular environment to trigger the oscillatory behavior, we acquire biochemical information from every single cell. To characterize the spatio-temporal behavior of the ensuing glycolytic synchronization, we then employ tools from graph theory. 
Finally, we test in a detailed kinetic model \cite{model1, single4, Dawie2019} the proposed mechanism of coupling between oscillatory cells leading to synchronization. We employ numerical simulations that integrate the kinetic model for the intracellular reaction network \cite{single4} with the physical geometry and hydrodynamic conditions in the experiment. These results are potentially useful to study the role of cellular synchronization in biomedical applications, such as insulin secretion regulation. 

\section*{Results}
\subsection*{Glycolytic oscillations in a microfluidic environment} 
\begin{figure}[b!]
\begin{center}
\resizebox{1.07\columnwidth}{!}
{\includegraphics{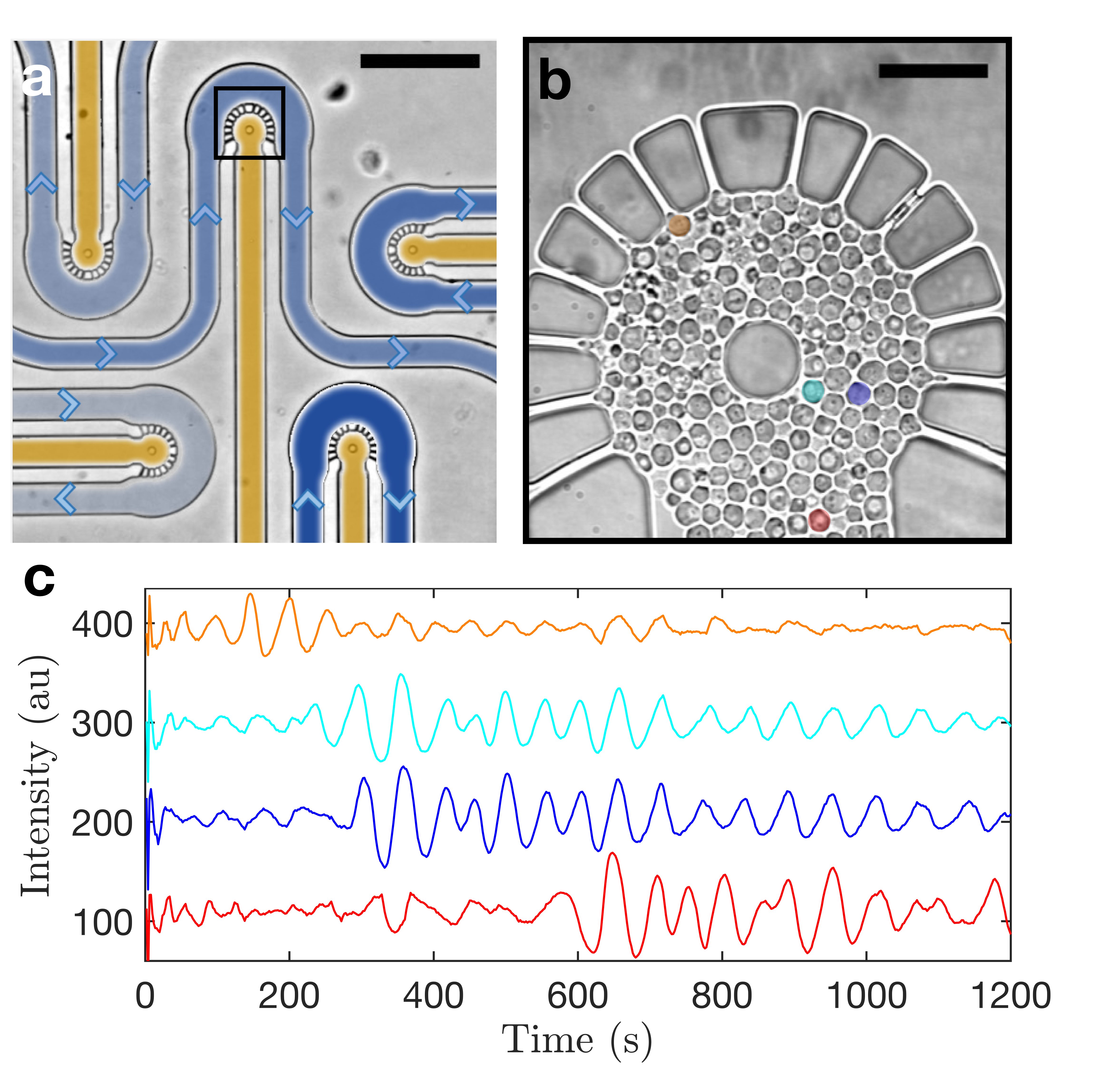}} 
\protect\caption{Glycolytic oscillations are affected by cell-position in the microfluidic chamber.
(a) Yeast cells from a single batch culture are loaded through the central yellow-shaded inlet channels of the diffusion chambers. They are then exposed to GLC and different concentrations of CN$^-$ (weighted blue shades) and NADH autofluorescence signals detected in individual cells. The length bar is 200~$\mu$m.
(b) Zoom-in of a loaded cell chamber showing the individual cells' locations. The length bar is 20~$\mu$m.
(c) Processed time series of the NADH concentration at the cells color-coded in (b) for the 12 mM CN$^-$ case. The intensity of the signals is shifted along the y-axis for visualization purposes. An NADH wave  propagates through the cells, starting at the top of the chamber (orange cell, from about 100~s), propagating to the middle of the (cyan and blue cells, from about 300~s), and finally to the bottom of the chamber (red cell, from about 500~s).} \label{fig:figureone}

\end{center}
\end{figure}

We designed a microfluidic environment that permits us to trigger and track glycolytic oscillations in an array of yeast cells (\textit{Saccharomyces cerevisiae}) with single cell resolution {(see methods sections ``Microfluidic device design and fabrication'' and ``Cell preparation'')}.
Figure~1(a) shows an image of the microfluidic chip with five chambers, one of which is highlighted by the black box and zoomed-in in Figure~1(b). 
The use of a single microfluidic device with multiple chambers permits us to load cells from a single batch and expose them to a range of stress solution concentrations in parallel, thus avoiding artifacts that might arise when performing experiments sequentially due to confounding factors such as cell storing time.
As shown in Figure 1(a), each cell chamber has an inlet channel (shaded in yellow), which we use to load the yeast cells into the chamber, and a perfusion channel (shaded in blue), which we use to expose the cells to a GLC and CN$^-$ solution with various concentrations of CN$^-$ (indicated by the different shades of blue). 
Importantly, the perfusion occurs by diffusion in quasi-static flow conditions through a series of diffusion apertures between the perfusion channel and the cell chamber, as can be seen in Figure~1(b). In this diffusion-limited cell chamber, the ACA produced by the cells is not washed away by convection and can mediate the cell-cell interactions.

Figure~1(b) zooms in on a loaded cell chamber where the single yeast cells can be clearly seen. 
We grow, harvest, and starve the yeast cells to obtain a strong oscillatory behaviour in response to GLC addition {(see methods section ``Cell preparation'')}. 
The cells are loaded at a controlled density in all cell chambers using a precision multi-syringe pump.
Afterwards, we inject the stress solutions containing  40~mM GLC with 8, 12, 16, 20 and 24~mM CN$^-$ using a second multi-syringe pump. We provide a constant stress solution supply for 20 minutes; during this time, GLC and CN$^-$ diffuse into the cell chambers through the diffusion apertures and are progressively consumed by the cells leading to the formation of a decreasing concentration gradient.
GLC consumption is linked to the production of reduced nicotinamide adenine dinucleotide (NADH), an intermediate metabolite which can be detected on an individual cell basis due to its autoflorescence. NADH show an excitation peak around a wavelength of 340~nm and emission peak around 460~nm {(see supplementary video 1)}. The details of the experiment are provided in the {Methods section ``Experimental procedure''} and those of the signal acquisition and analysis in the {Methods section ``Signal acquisition and conditioning''}.

The NADH signals of four different cells are shown in Figure~1(c) for the 12~mM CN$^-$ case. They are color-coded to correspond to the cells highlighted in Figure~1(b).
Depending on their positions in the chamber, the cells start to oscillate at different times: the orange cell is located next to a diffusion aperture and about 100~s after the stress solution injection it starts to show oscillations with a transient increase in the amplitude; the cyan and blue cells are located further away from the diffusion apertures; and they display sustained oscillations and the amplitude increase at about 300~s; the red cell is even further away from the diffusion apertures, and the strong oscillations appear only after about 600~s.
The  amplitudes of these NADH signals depend on the CN$^-$ concentration \cite{single4, single5, yeast2}, which follows the diffusion gradient away from the diffusion apertures:
 the amplitude of the orange cell signal is smaller than those of the cyan and blue cells, which in turn are smaller than that of the red cell.

Finally, local synchronization between cells occurs because of their secretion and exchange of ACA, which determines the local cell-cell coupling \cite{single4, model1}. For example, the cyan and blue cells are close to each other and, thus, their signals are highly synchronized for the whole duration of the experiment.

\subsection*{Synchronization analysis} 

\begin{figure}[b!]
\begin{center}
\resizebox{1\columnwidth}{!}
{\includegraphics{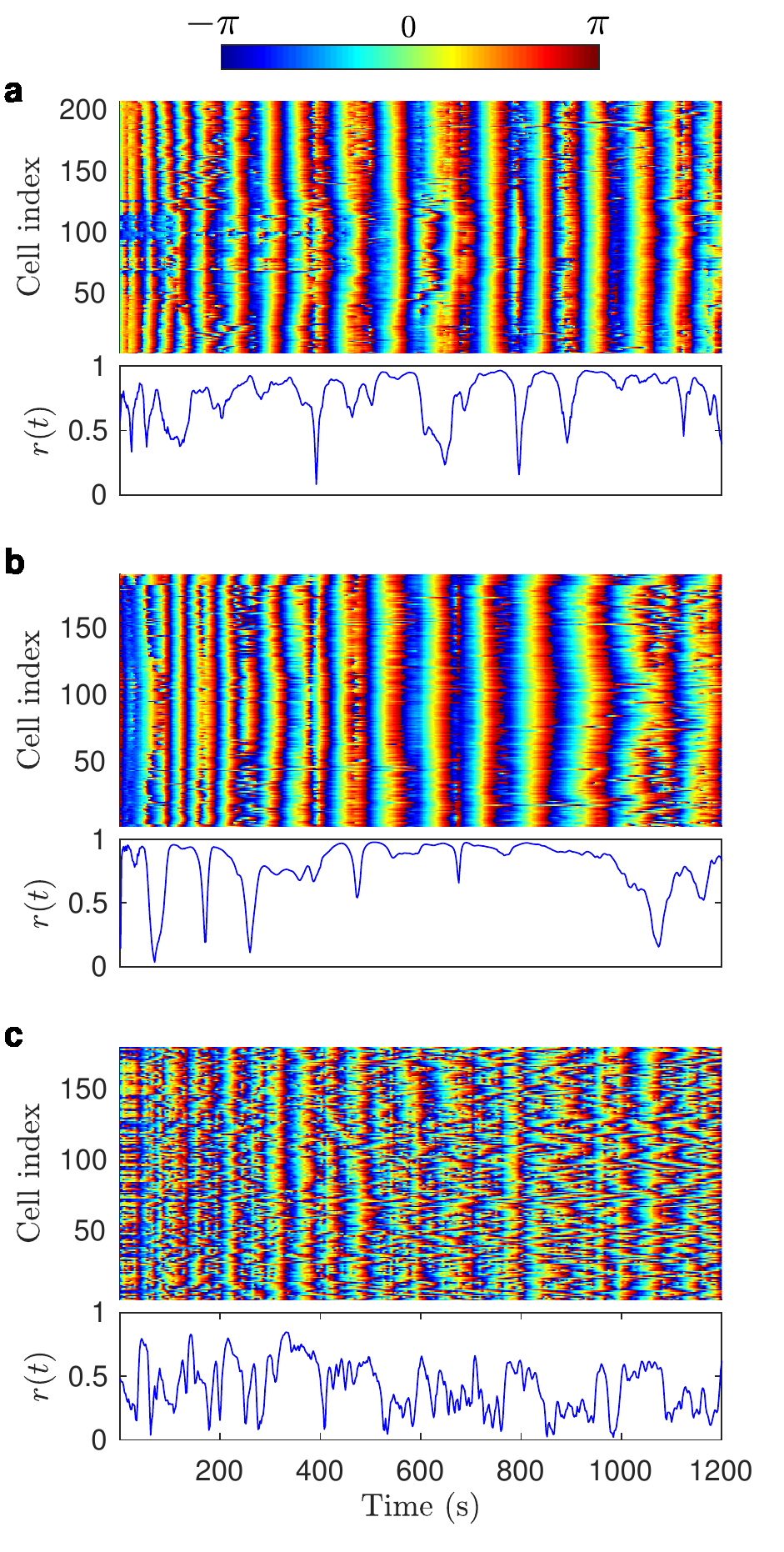}} \protect\caption{Coupling between the glycolytic signals of the single cells. 
Instantaneous phases (-$\pi $ to $\pi $) of the NADH autofluorescent signals for all cells (top panels) and normalized order parameter $r(t)$ (bottom panels) when glycolytic oscillations in yeast cells are triggered using (a) 12~mM, (b) 16~mM and (c) 20~mM CN$^-$ combined with 40~mM GLC.
(a-b) At intermediate CN$^-$ concentrations, the instantaneous phases feature global patterns across the cell array corresponding to the periods when $r(t) \approx 1$ maxima of the order parameter, suggestive of a spatio-temporal synchronization.
(c) At high CN$^-$ concentrations, this synchronization is lost, reflected in the low  values of the order parameter.
} \label{fig:figurethreea}
\end{center}
\end{figure}

For the synchronization analysis of the NADH signals, we extract instantaneous phases of the discrete Hilbert transforms of the time series. (color-coded phase plots in Figure~2). From these phases, we calculate the time-dependent order parameter $r(t)$ normalized between 0 and 1, which measures the overall degree of synchrony of the cell array (lower plots in {Figure}~2) \cite{kuramoto2} {(see methods section ``Synchronization analysis'')}.

\begin{figure*}[t]
\hspace*{-1.3 cm}
\resizebox{2.4\columnwidth}{!}
{\includegraphics{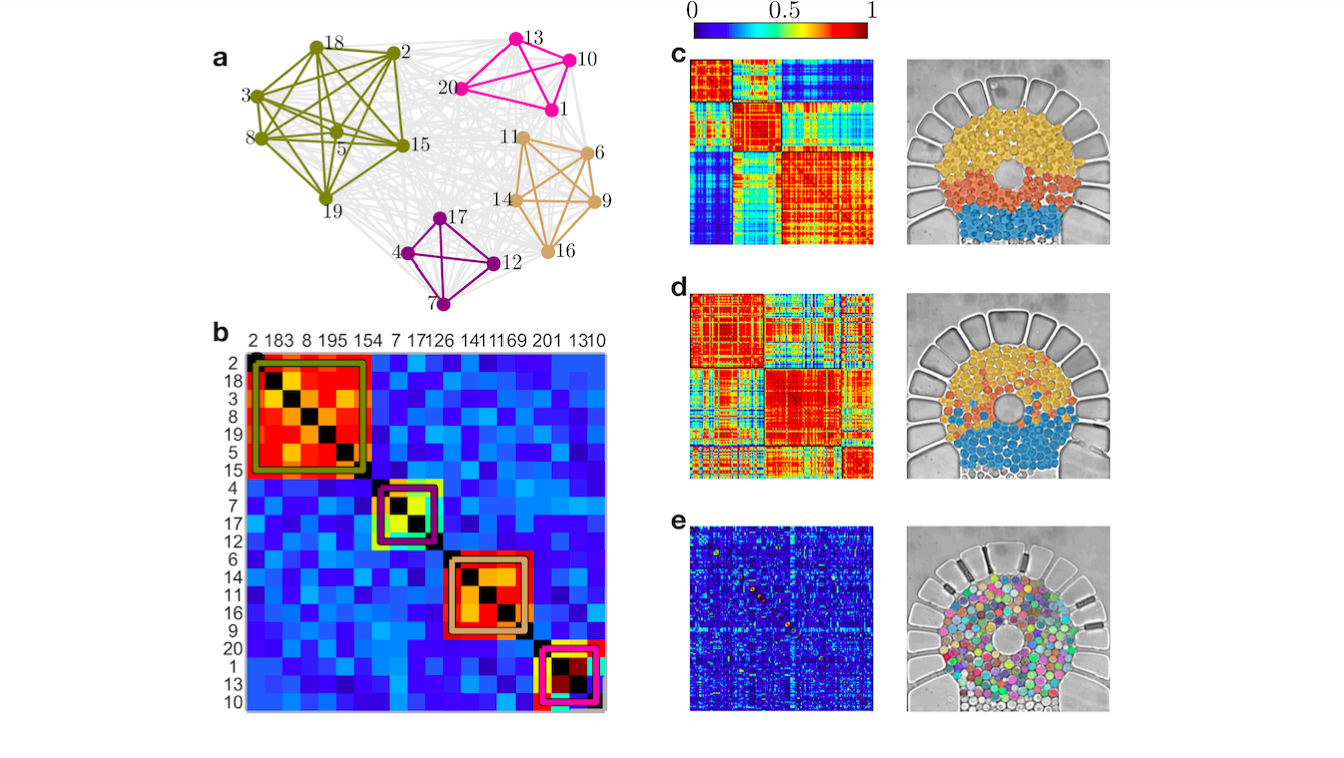}} \protect\caption{ Synchronization communities. 
(a) From the NADH autofluorescent signals of the cells, we construct a graph where each cell is a node, and the correlation between the signals of two cells is a measure of their connection strength. 
In the resulting graph, it is possible to identify communities of cells (colored subgraphs) that are well connected with each other, but poorly connected to cells belonging to different communities.
(b) Such community structure is reflected in the adjacency matrix representing the graph. Note that the order of the nodes has been rearranged to more clearly highlight the community structure.
(c-e) Adjacency matrices and corresponding communities overlaid on the images of the corresponding cell arrays for (c) 12~mM, (d) 16~mM and (e) 20~mM CN$^-$. 
}\label{fig:figurefour}
\end{figure*}

When the concentration of CN$^-$ in the stress solution is sufficiently high (12~mM, Figure~2(a), and 16~mM, Figure~2(b)), the majority of the cells exhibit sustained oscillatory behaviour and synchronization, which is shown by the fact that $r(t)$ is consistently very close to 1.
In both cases, $r(t)$ features some minima, which reflect temporary incoherent behavior between the oscillating cells. 

When the CN$^-$ is even higher (20~mM CN$^-$, Figure~2(c)), the cells still oscillate but the phase plot does not show any global synchronization between them, which is reflected in the fact that $r(t)$ is consistently smaller than 1 and fluctuates for the whole duration of the experiment.
This suggests the absence of local coupling between cells, that the ACA lowering due to binding by cyanide weakens the local coupling to such an extent that it cannot lead to global synchronization. The cases for 8~mM and 24~mM CN$^-$ are shown in {Supplementary Figure~2.}

\subsection*{Synchronization communities}

\begin{figure*}[t]
\hspace*{-1.2 cm}
\resizebox{2.14\columnwidth}{!}
{\includegraphics{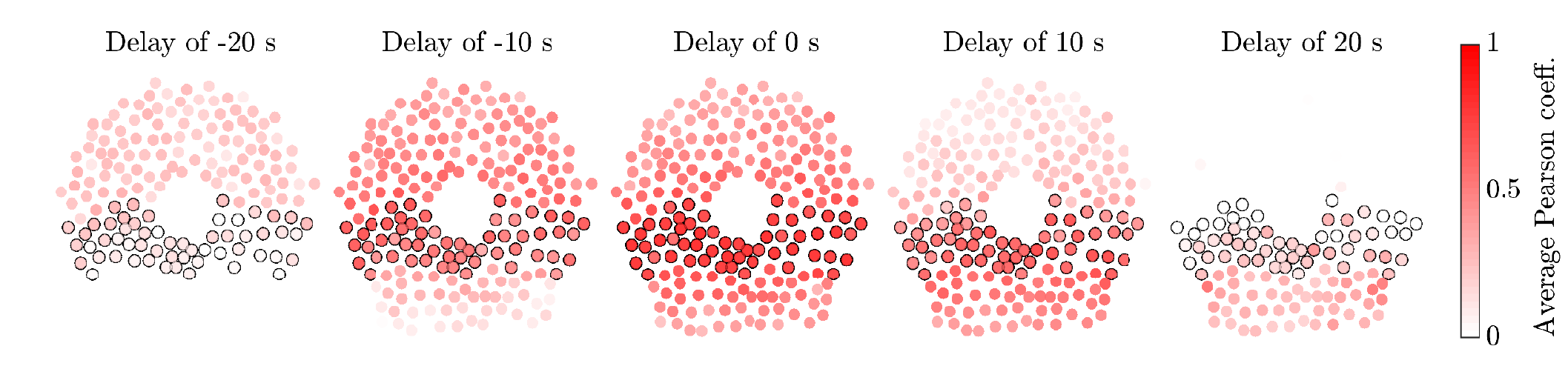}} \protect\caption{ Propagation of a glycolitic synchronization wave.  
Average delayed correlations between a community of synchronized cells (marked with black rings) and every cell present in the chamber (red circles) for a cell array exposed to 12~mM CN$^-$.
The delayed Pearson correlations are calculated with delays from -20~s to +20~s and feature a series of maxima that propagate from top to bottom.
The shape of this glycolytic wave depends on the chemical diffusion profiles of stress solution and secreted metabolites, and it is influenced by the geometrical constraints of the device.} \label{fig:figurefive}
\end{figure*}

To test for the existence of locally synchronized communities within the population, we make use of graph theory and community analysis  {(see methods section ``Graph construction and community analysis'')}.
Using a technique that is commonly employed in studying connectivity between brain regions \cite{graph2}, we determine the coupling strength between cells based on the degree of synchrony in their glycolytic signals.
The resulting graph can be represented as a series of nodes (cells) and edges (correlation) (Figure~3(a)) or, more conveniently for analysis purposes, as an adjacency matrix (Figure~3(b)), which is a square matrix where each entry represents the strength of the connection between the nodes corresponding to its row and column indices. 

A graph can be divided into communities so that the nodes within each community are more strongly connected with each other than with the rest of the graph. 
We identify the communities using the Louvain algorithm \cite{graph0}. 
For example, we show the color-coded communities on the schematic of the graph in Figure~3(a) and on the adjacency matrix in Figure~3(b).

Figures 3(c-e) show the adjacency matrices and the corresponding communities overlaid on the cell images for cell arrays exposed to different CN$^-$ concentrations. 
When exposed to 12~mM CN$^-$ (Figure~3(c)), three well-defined communities form. The community in yellow covering half of the circular chamber is radially exposed to the stress solution through the diffusion apertures. Hence, these cells experience simultaneous triggering of the oscillations. The community in red is triggered by the resulting concentration of the stress solution that passes through the yellow community and the lateral diffusion apertures. Consequently, the oscillations appear with a delay in respect to those of the yellow community. Finally, the cells in blue are not directly exposed to the diffusion apertures. This community shows the longest delay and will be exposed to lower concentrations of GLC and CN$^-$. Each of the communities remains synchronized due to the local exchange of ACA. 

When the CN$^-$ concentration is increased to 16~mM (Figure~3(d)), the boundaries between the communities become less defined. 
The yellow community (top half ) forms similarly to the 12~mM scenario. However, a second community (red) appears with scattered cells at different locations in the top half of the cell chamber. This can be explained by the mentioned uncoupling effect that higher concentrations of CN$^-$ can produce between neighboring cells, despite the fact that the oscillations are triggered simultaneously. With a less defined boundary, a third community (blue) appears where cells are not exposed to diffusion apertures. 
Interestingly, one of the diffusion apertures in this specific cell chamber (bottom-right in Figure~3(d)) is obstructed, and this deforms the resulting  community, confirming its dependence on the direct diffusion of the stress solution.
At 20~mM CN$^-$ (Figure~3(e)) consumption of ACA is so high that the cells display uncorrelated behavior, and the community structure breaks down.

\subsection*{Glycolytic synchronization waves}

{Supplementary video~1} shows the NADH autofluorescence spatio-temporal distribution across the cell array. We characterize the arising glycolytic wave by calculating the delayed cross-correlation between the NADH signals in a community of cells identified with the modularity analysis (black rings in figure~4), and the rest of the cell array {(see methods section ``Graph construction and community analysis'')}. The wavefront propagation is then displayed as a relative measure with respect to one of the communities obtained at zero delay. Hence, the gradual increase in each cell correlation level translates into a reduction of the phase difference between the cell and the reference community. The obtained wave therefore describes the transition between the communities as a function of their relative delays, giving a phase relation between them.

Figure~4 shows the spatio-temporal evolution of a synchronization wave propagating in a cell array exposed to 12~mM CN$^-$. The high values of the normalized cross-correlations (dark red) travel from the region mostly exposed to the diffusion apertures, to the region containing the cells further away from the direct exposure to the stress solution. It can be noticed that at zero delay the cells in the reference community (black rings) give an approximate profile of the propagating wavefront. Due to the cell heterogeneity and the discrete nature of the cell monolayer, local cell-cell interactions show small variations in the wavefront shape. However the overall synchronization wave can be tracked despite these deviations.

\subsection*{Simulations}

We simulate a 2D array of cells with the same structure as the experimental one. We calculate the time-dependent concentrations of GLC, CN$^-$, ACA and ethanol (ETOH) {(see supplementary video 2)} together with all the intracellular metabolites. 

To test the ACA coupling, we used a previously published detailed enzyme mechanistic model \cite{single4} to simulate glycolysis in each cell as a set of ordinary differential equations. Transport and diffusion of extracellular metabolites were simulated using partial differential equations with the physical characteristics of the microfluidic chamber determining the boundary conditions. {(see methods section ``Numerical simulations'')}. 
We then extract the glycolytic signals (NADH concentration) for each cell and analyze them as in the experimental data (synchronization analysis and community structure analysis).
{Supplementary Figure 1} shows the NADH instantaneous phase distribution and the community structure for the simulated cases of 20, 24 and 28~mM~CN$^-$. 
As the CN$^-$ concentration is increased, the communities show less defined boundaries, in good agreement with the experiments (Figure 3), albeit that higher cyanide concentrations were used in the simulations.
High CN$^-$ concentrations only induce transient oscillations leading to a steady state (Supplementary Figures 1(b-c)).

The simulated time-dependent ACA distribution for external and internal concentrations {(see supplementary video 2)} reflect an adaptation to travelling waves. The initial oscillations display uncorrelated behaviour that gradually transforms to periodic propagations across the cell array. The simulation results are in good agreement with the experimental results although wavefronts in the simulations show a more continuous profile than is observed in the experimental data. This is most likely due to an incomplete capture of cellular heterogeneity in the model.

 The high external GLC concentration supply (40 mM) induces a fast diffusion until homogeneous covering of the complete monolayer {(see supplementary video 3)}. In contrast, the diffusion of CN$^-$ shows a quasi-constant distribution profile to which the ACA wavefront shapes adapt. The periodic fluctuations given by the external ACA and CN$^-$ reaction show negligible influence on the CN$^-$ levels as it has a much higher absolute concentration.

\section*{Discussion}

We presented a three part methodology to study the coupling between individually oscillating cells in yeast at a single-cell level. First we detected the single-cell metabolic responses from yeast cells in a microfluidic device. Second, we identify synchronized communities and track the synchronization waves using graph theory. Third, the underlying mechanism for cell-cell communication was tested in a validated mechanistic model for individually oscillating yeast cells, where each glycolytic network is linked to the architecture and physics of the microfluidic system.

We have shown that the lateral metabolic coupling between individual cells can induce metabolic synchronization waves.
The implemented microfluidic device permits us to control the concentration of extracellular chemicals required to trigger glycolytic oscillations, ensuring a constant supply of GLC and CN$^-$ by direct diffusion. 
After addition of the stress solution, the start of oscillations in individual cells is dependent on the position of cells in the cell chamber.
The spatial distribution of CN$^-$ concentration influences the overall degree of synchronization, which is reflected in the time-evolution of the extracted instantaneous phases and order parameter.  
Furthermore, even in the presence of low overall values of synchronization, there can be communities of cells that are synchronized at the local level, which we have identified using graph theory analysis.

Our findings corroborate the observations of previous studies on macroscopic glycolytic waves \cite{waves_extract, waves1, waves2} and explain these waves in terms of metabolic coupling between individually oscillating cells.
In \cite{Dawie2019} it was demonstrated, through experimental observation and analyses using a detailed mechanistic model, that single oscillating cells can adapt their phases to external ACA signals. These signals are transduced through the system of cofactors which regulates the activity of phosphofructokinase and leads to the correct movement of the phase to allow for synchronization. Here we show that the spatio-temporal dynamics of synchronization waves in a population are well-described by the same detailed mechanistic model (constrained to the physical conditions of the cells in the microfluidic chamber) showing that the proposed ACA coupling between the cells is sufficient for the observed collective behaviour.

The method we developed allows for the combination of detailed mechanistic models at single cell level, with the experimental analyses of collective synchronized responses at the population level. Such an approach can also be used for studying the spatio-temporal dynamics present during insulin production by pancreatic $\beta$ cells \cite{beta0, beta1, beta2, beta3}. Furthermore, the presented analysis can be applicable to other biological systems that display synchronization of individual oscillators e.g. wave propagation in the heart leading to muscle contraction and synchronized oscillatory phenomena in groups of neurons.

\section*{Methods}

\subsection*{Microfluidic device design and fabrication} 

\noindent Each chamber where the cells are loaded (diameter 55~$\mu$m, height 5~$\mu$m, Figure~1(b)) is surrounded by a series of diffusion apertures (width 2~$\mu$m). These diffusion apertures are connected to a perfusion channel (width 65~$\mu$m, weighted blue shades in Figure~1(a)), where the stress solution flows through inlet and outlet channels (width 50~$\mu$m). A cylindrical pillar (diameter 3~$\mu$m) is placed at the center of the chamber to prevent the chamber ceiling from bending.

Silicon molding masters are fabricated using photolithography. A negative photoresist (SU-8 3005, MicroChem Corp., Newton, MA, USA) is spin-coated (3500~rpm, 30~s), soft-baked (2~minutes at 65$^\circ$C; 3~minutes at 95$^\circ$C), UV-exposed (15~mW~cm$^{-2}$ for 10~s (Suss MicroTec SE., Garching, Germany) under HardContact pressure mode), post-exposure-baked (3~minutes at 65$^\circ$C; 4~minutes at 95$^\circ$C), and developed (2~minutes, SU-8 developer mr-Dev 600, Micro resist technology GmbH).

For the molding procedure, we have followed the established procedure described in Ref.~\cite{PDMSfirst}. Briefly, polydimethylsiloxane (PDMS) is homogeneously mixed with a curing agent (Sylgard 184 Silicone Elastomer Kit, Dow Corning Corp. Seneffe, Belgium) in a 15:1 ratio. The mixture is degassed using a vacuum dessicator (30~minutes), poured onto the master, and baked (3~hours at 90$^\circ$C). The resulting PDMS structure is covalently bond to a cover glass (thickness No. 1 (0.13 to 0.16~mm), 45$\times$60~mm, HECH990/6045, VWR) using oxygen plasma (40~s, PDC-32G, Harric Plasma).

\subsection*{Cell preparation} 

\noindent The yeast cell strain used in the experiments is X2180 \textit{Saccharomyces cerevisiae}. Single colonies are grown following the same protocol used in Refs.~\cite{single1, cellprep}. 
The cells are grown in a carbon source medium containing 10~g~L$^{-1}$ GLC, 6.7~g~L$^{-1}$ yeast nitrogen base (YNB), and 100~mM of potasssium phthalate at pH~5. The suspensions are cultured in a rotary shaker at 30$^\circ$C until GLC depletion in the media. 
To achieve the diauxic shift (GLC starved and switched to a slower exponential growth), the cells are washed and starved in 100~mM potassium phosphate (pH~6.8) for 3 more hours in the rotary shaker at 30$^\circ$C. 
Finally, in order to maintain the cells in the diauxic shift, they are washed and stored at 4$^\circ$C until the experiments. 

\subsection*{Experimental procedure}

\noindent Cells are loaded into the five cell chambers using 250~$\mu$L glass syringes connected via polytetrafluoroethylene tubing (inner diameter 0.012~in, Cole-Parmer, Vernon Hills, IL, USA). 
In order to obtain equal cell densities, the cell solution is introduced via equal length microfluidic paths and flow rates (40 nL/min ) until the cell chambers are filled. After cell loading, the experiment is initiated with the injection of 40~mM GLC and 8~mM, 12~mM, 16~mM, and 24~mM CN$^-$ stress solutions, which circulate in the perfusion channel surrounding around each cell chamber at 25~nL~min$^{-1}$. This perfusion flow remains constant for the 20 minutes corresponding to the complete experimental acquisition. The cell loading and stress injection are performed using a precision multi syringe pump (CMA 400, CMA Microdialysis, Solna, Sweden).  

\subsection*{Signal acquisition and conditioning}

\noindent Image acquisition is performed using an inverted microscope (DMi 6000B; Leica Microsystems, Wetzlar, Germany) with a 100$\times$, NA=1.33 oil-immersion objective in an epifluorescence configuration. 
In order to measure the NADH autofluorescence intensity fluctuations, a 350/54 excitation filter and a 415/64 emission filter (DAPI set) are used together with a 15~W mercury short-arc reflector lamp (EL6000, Leica Microsystems, Wetzlar, Germany) \cite{NADHpeaks}. An EMCCD camera (C9100-12, Hamamatsu Photonics, Shizuoka, Japan) is used with an exposure time of 400~ms. Images are acquired every 2~s for a total period of 20~minutes using an automatized illumination, positioning, and acquisition routine programed using OpenLab (PerkinElmer, Waltham, MA, USA). 

The time series for the individual cells are obtained from the NADH autofluorescence images. 
For each cell and frame, the average intensity $x_n(t)$ of cell $n$ is computed over the region of interest (ROI) corresponding to the cell area. 
Using MATLAB$^{\tiny\textregistered}$, a background signal and a running averages of 55 data points are subtracted from the signal to reduce noise and short-term fluctuations. 

\subsection*{Synchronization analysis}

\noindent Starting from the glycolythic signals $x_n(t)$, the phase of cell $n$ is calculated as 
\begin{equation}\label{eq:phase}  
\Phi_n(t)={\rm arctan}
\left[ 
\frac{H(x_n)(t)}{x_n(t)}
\right], 
\end{equation}
where $H(x_n)(t)$ is the Hilbert transform of $x_n(t)$ evaluated with the MATLAB$^{\tiny\textregistered}$ built-in function.
These data are shown in Figure 2 and Supplementary Figure~1(a-c) for the experimental and simulated signals, respectively.

To evaluate synchronization, as standardized in previous works \cite{synchyeast1, single5}, the time-dependent order parameter $r(t)$ is obtained from the expression:
\begin{equation}\label{eq:orderp} 
r(t)= \left | \frac{1}{N}\sum^{N}_{n=1}e^{-i\Phi_{n}(t)} \right |,
\setlength{\abovedisplayskip}{10pt}
\setlength{\belowdisplayskip}{10pt}
\end{equation}
where $N$ is the number of cells in the cell chamber and $\textbf{ $\Phi $}_{n}$  is the instant phase for each yeast cell. 
The order parameter is normalized between 0 and 1.
When $r(t)$ is large, the individual cells' phases are synchronized; when $r(t)$ is small, there is high heterogeneity in the individual cell phases \cite{kuramoto0, kuramoto1}.
The degree of synchrony is characterized with $r$ values from 0 to 1 where low order parameter translates into high heterogeneity in the instant phases.  

\subsection*{Graph construction and community analysis}

By using graph theory, all the oscillating cells are considered as nodes of a network with connections weighted by the correlation of their signals. Synchronization distribution can then be characterized in terms of the formation of cell communities showing higher coherence in their signals.

The community structure algorithms aim to optimize the modularity, a measure of the quality of the community division of the network. In short, modularity measures the density of the connections within a community and compares it with what it would be in a given random network. The more positive modularity indicates a better division of the network into communities. The Louvain algorithm approaches the problem of modularity maximization by iteratively grouping single nodes into communities. It starts by assigning each node in the network to a separate community. By changing the community participation of a node and its neighbors, it optimizes the modularity locally throughout the network. This results in having some community structure in the network. In the second step, these communities become nodes, and the first step of local modularity maximization is reapplied. These two steps are repeated until the maximum modularity is obtained and there are no changes in modularity values with any new iteration. Finally, the community organization of the step with maximal modularity is taken to be the real and final community of the network.

By setting a threshold in the correlation
coefficient of 0.7, subgroups of cells showing synchronized behavior are obtained for the different concentration ratios in the stress solution. The functions to perform this process are adapted from the MatLab-based software BRAPH \cite{graph2}. A correlation adjacency matrix weighted with the correlation coefficients can be constructed by rearranging the node indices in subgroups showing higher connectivity. The indices assigned to each community can then be mapped based on their original location in the cell array and display the community spatial distribution.

To map the spatial distribution of the phase in form of wavefronts propagating across the cell array, the average correlation coefficient is calculated between each cell signal and a reference synchronization community at different delays.

\subsection*{Numerical simulations}

The numerical simulations combine a kinetic model for single-cell glycolysis  \cite{model1} with the geometrical and hydrodynamical conditions given by the cell arrangement in the microfluidic chip. The flow velocity field and the concentration gradients through the device are calculated using the finite-element based interpolation software COMSOL Multiphysics (COMSOL Inc., Burlington, MA, USA). The device geometry is defined with no-slip boundary conditions, and the nodes for the numerical interpolation are generated using the "Extra fine", Physics controlled mesh mode. The fluid inside the device design is considered as Newtonian and incompressible, which obeys the Navier-Stokes equation for the stationary case:
\begin{equation}\label{eq:navier} 
\rho(\textbf{u}\cdot\nabla)\textbf{u} = 
-\nabla p+\eta\nabla^{2}\textbf{u}+\textbf{f} 
\setlength{\abovedisplayskip}{12pt}
\setlength{\belowdisplayskip}{12pt}
\end{equation}
Where \textbf{u} is the flow velocity, \textbf{f} represents body force densities \textendash which are negligible for this case\textendash and the constants $\rho$ and $\eta$ are the density and dynamic viscosity respectively corresponding to water at room temperature; finally, \textit{p} represents the pressure \cite{microfluid}. The density and dynamic viscosity of the fluid is considered as the predefined for water by the software.
To simulate the chemical transport and distribution, the time-dependent concentration gradients are calculated in order to follow diffusion across the cell chamber. The relation describing the process is given by the convection-diffusion equation: 
\begin{equation}\label{eq:convective} 
\frac{\delta c}{\delta t}=- \nabla\cdot(-D \nabla c+c \textbf{u})
\setlength{\abovedisplayskip}{10pt}
\setlength{\belowdisplayskip}{10pt}
\end{equation}
Here, $c$ is the concentration and $D$ is the diffusion coefficient. The stress solution contained GLC and CN$^-$  with diffusion coefficients of 6.7x10$^{-10}$ m$^{2}$/s and 20.7x10$^{-10}$ m$^{2}$/s, respectively. In addition, the concentration gradients are calculated for the time-dependent secretions of ETOH and ACA from the cells, with diffusion coefficients of 1.15x10$^{-9}$ m$^{2}$/s and 1.3x10$^{-9}$ m$^{2}$/s respectively. Initial concentrations in the chamber are defined with the minimum values required to guarantee the oscillatory state of all the cells \cite{single1}, with 5 mM CN$^-$ , 9 mM GLC, 1x10$^{-6}$ mM ACA and 1x10$^{-6}$ mM ETOH. 

A total of 210 circular boundaries are defined with sizes in the range of yeast cells at different ages \textendash between 4 $\mu$m and 10 $\mu$m \textendash and are distributed inside the chamber design in a dense monolayer to emulate the experimental conditions. Global definitions of the membrane diffusion coefficients are assigned for ETOH, ACA, and CN$^-$  with values of 5.88x10$^{-12}$ m$^{2}$/s, 5.87x10$^{-12}$ m$^{2}$/s and 5.88x10$^{-12}$ m$^{2}$/s respectively. On the other hand, Diffusion coefficients inside the cells are defined high enough (1 m$^{2}$/s) to consider the chemical concentrations to be homogenous and obtain single values for each cell geometry. The kinetic model previously described for individual cells  in a microfluidic device \cite{single1, single3, model1, Dawie2019} is implemented to calculate the metabolite concentrations inside the cells as well as the secretions of ACA and ETOH in the chamber. For these calculations, the GLC and CN$^-$  total exposure is defined by the convection-diffusion equation in combination with the lactonitrile formation from the reaction of CN$^-$ with ACA \cite{ACA1, lacto}. The initial concentrations inside the cells are defined with a global initial intracellular GLC level of 3 mM and a heterogeneity of 10 different values of initial CN$^-$ , ACA and ETOH. Similarly, 10 different initial values are assigned for all the metabolites taking place in the ordinary differential equations along the metabolic pathway. The reaction rates taking place inside the cells are defined for each of the circular domains for GLC, CN$^-$ , ACA, and ETOH (for details on all the rate equations see the interactive online model applicative in https://jjj.bio.vu.nl/models/gustavsson5/simulate/). The resulting concentrations are tracked in time by individual probes inside each cell and in the extracellular media.


\beginsupplement
\begin{figure*}[t!]
\begin{center}
\hspace*{-1.2 cm}
\resizebox{2.4\columnwidth}{!}
{\includegraphics{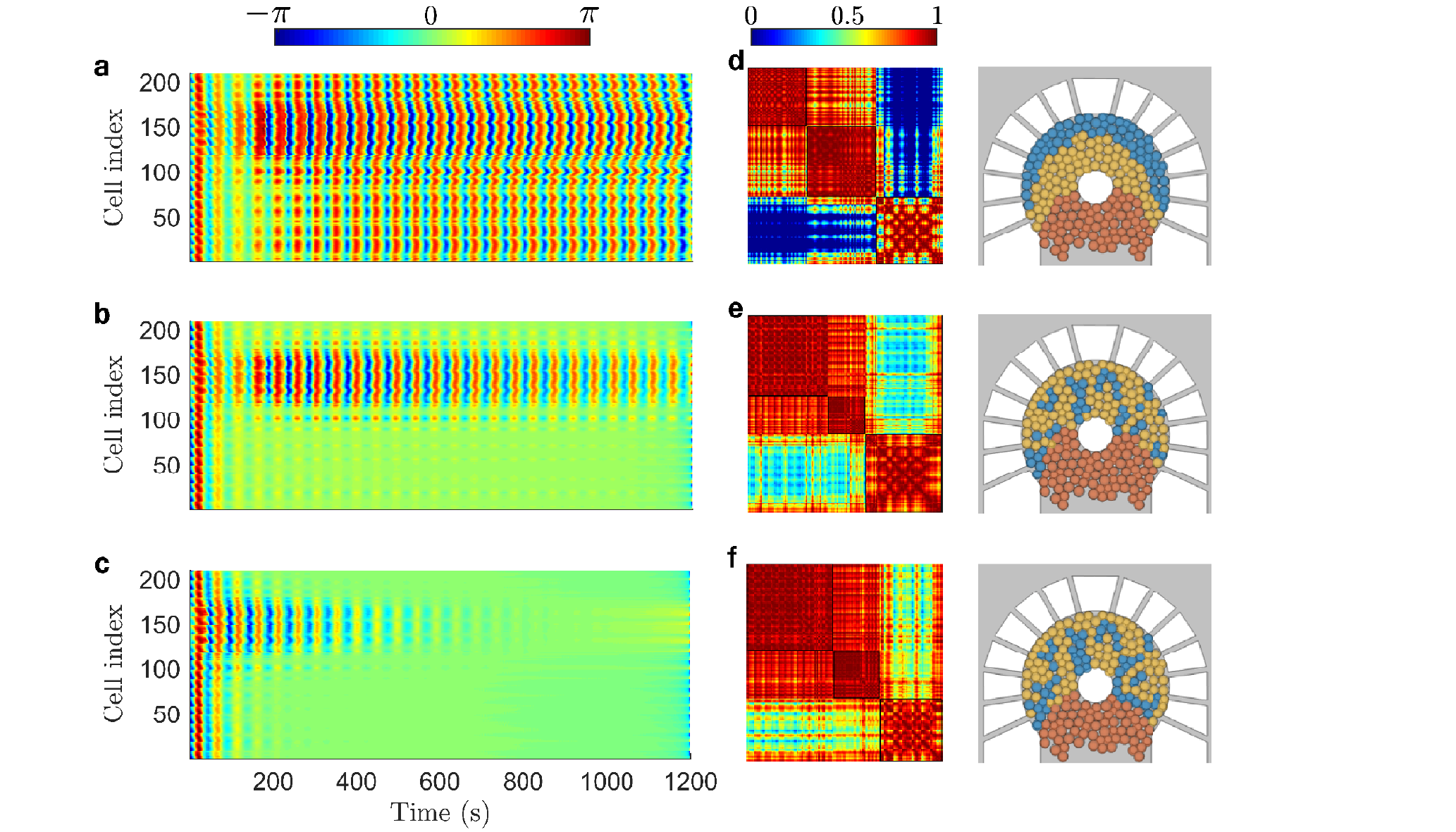}} \protect\caption{ The kinetic model for single cells predicts the formation of synchronization communities. The 210 simulated cells using the Gustavsson model under the boundary conditions that resembled the experiments, display coupled oscillatory behaviour in the metabolites present through the individual glycolytic pathways. For the  CN$^-$ concentrations in the stress solution of (a) 20 mM, (b) 24 mM and (c) 28 mM, NADH instantaneous phases (from -$\pi $ to $\pi $) showed the most distinguishable cases of synchronization distribution across the cell array. (d)- (f) The weighted matrices with Pearson correlation coefficients from 0 to 1 underline the synchronization communities, that emerge shaped by the diffusion gradients from the geometrical conditions. In contrast with the experimental results, a steady state was achieved more homogeneusly as cells were exposed to the higher CN$^-$ concentrations, resulting into new non-oscillating communities. Community colors are assigned randomly.} \label{Supplementary figure1}
\end{center}
\end{figure*}

\begin{figure*}[t!]
\begin{center}
\hspace*{-1.2 cm}
\resizebox{2.4\columnwidth}{!}
{\includegraphics{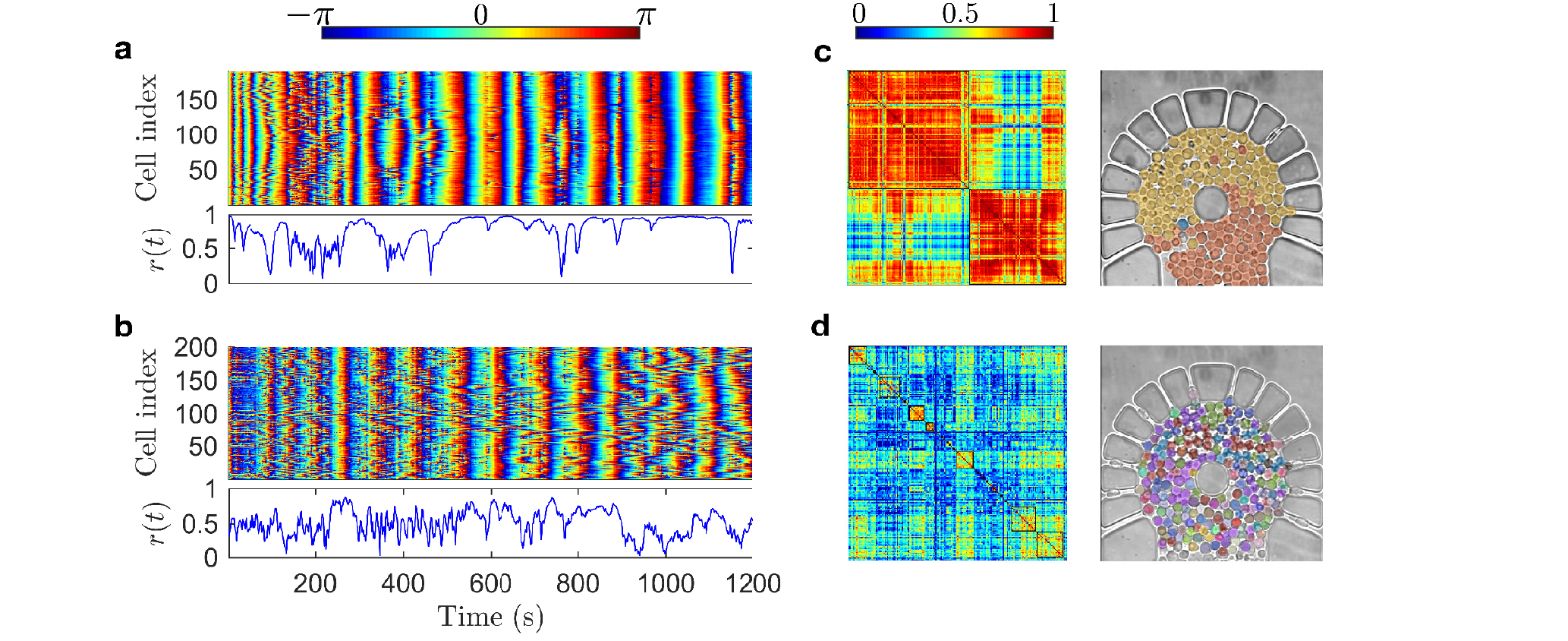}} \protect\caption{ Synchronization analysis and community structure for the experimental cases of 8~mM and 24~mM of CN$^-$. (a) For concentrations sufficiently low, glycolytic oscillations are less sustained. Despite the fact that the order parameter $r$ can show higher values, oscillations not necessarily correspond to the glycolytic cycle. (b) Significantly high concentrations of CN$^-$ induce uncorrelated behaviour among the cells as can be noticed from the low values of the order parameter $r$. (c-d) Present the community structures for these two extreme experimental cases, where the 8~mM case reveal larger communities due to the high presence of secreted ACA. On the other hand in the 24~mM case, a higher amount of ACA is consumed by the binding with CN$^-$ resulting in an undefined community structure.  Community colors are assigned randomly.} \label{Supplementary figure2}
\end{center}
\end{figure*}

\end{document}